\documentclass[12pt]{iopart}
\usepackage{graphicx}
\usepackage{dcolumn}
\usepackage{bm}
\usepackage{color}
\usepackage{cite}
\usepackage{hyperref}%
\hypersetup{
    colorlinks=true,
    linkcolor=blue,
    citecolor=blue,   
   urlcolor=blue,
   }

\def\bea{\begin{eqnarray}}
\def\eea{\end{eqnarray}}
\def\beq{\begin{equation}}
\def\eeq{\end{equation}}

\def\be{\beta}

\def\t{\tau}

\def\Pe{{\rm Pe}}

\def\la{\langle}
\def\ra{\rangle}
\def\nn{\nonumber}

\def\S{\Sigma}

\def\d{\delta}
\def\p{\partial}
\def\l{\Pe}

\def\w{\omega}

\def\uv{ \hat{\bm{u}}}
\def\rv{ {\mathbf r}}
\def\vv{ {\mathbf v}}

\def\d{\delta}
\def\p{\partial} 

\def\la{\langle}
\def\ra{\rangle}

\begin{document}
\title[Inertial dynamics of ABPs]
{Exact moments for trapped active particles: inertial impact on  steady-state properties and re-entrance}

\author{Manish Patel}
\address{Institute of Physics, Sachivalaya Marg, Bhubaneswar 751005, India}
\address{Homi Bhabha National Institute, Anushaktinagar, Mumbai 400094, India}
\ead{manish.patel@iopb.res.in}

\author{Debasish Chaudhuri}
\address{Institute of Physics, Sachivalaya Marg, Bhubaneswar 751005, India}
\address{Homi Bhabha National Institute, Anushaktinagar, Mumbai 400094, India}
\ead{debc@iopb.res.in}

\begin{abstract}
In this study, we investigate the behavior of inertial active Brownian particles in a $d$-dimensional harmonic trap in the presence of translational diffusion. While the solution of the Fokker-Planck equation is generally challenging, it can be utilized to compute the exact time evolution of all time-dependent dynamical moments using a Laplace transform approach. We present the explicit form for several moments of position and velocity in $d$-dimensions. 
An interplay of time scales assures that the effective diffusivity and steady-state kinetic temperature depend on both inertia and trap strength, unlike passive systems.  We present detailed `phase diagrams' using kurtosis of velocity and position showing possibilities of re-entrance. 
\end{abstract}


\maketitle

\section{\label{sec:introduction}Introduction: }
 Active matter comprises microscopic entities exhibiting autonomous motion, propelled by harnessing either ambient or internal energy at the smallest scale, thereby maintaining a state of non-equilibrium and violating the equilibrium fluctuation-dissipation relation~\cite{Bechinger2016, Marchetti2013,Romanczuk2012,Ramaswamy2019}.
In nature, these self-propelling entities manifest across various scales, spanning from motor proteins, cells, and tissues to insects, birds, and animals~\cite{brownianmotor02,Reimann2002,Berg1972,Niwa1994,Ginelli2015,Devereux2021,Mukundarajan2016,Rabault2019}.
Active entities performing orientationally persistent random motion are termed active Brownian particles (ABPs).
Other related models, such as run-and-tumble particles (RTPs) and the active Ornstein-Uhlenbeck process (AOUP), exhibit similar dynamical properties up to second-order moments~\cite{Cates2013, tailleurnoneq, Das2018a, Shee_20}.
These models are extensively investigated under the assumption of overdamped dynamics, as seen in active Janus colloids with an extremely short inertial relaxation time of around $\sim 100$\,ns compared to their relevant persistence time $\sim 1-10$\,s~\cite{Kurzthaler_18}. 
However, larger active entities like insects, birds, animals, and macroscopic artificial active matter such as vibrated rods, granular particles, active spinners, hexbugs, and vibrobots possess significant mass. They may exhibit much slower inertial relaxation~\cite{Scholz_2018,Narayan2007,Kudrolli2008,Deseigne2010,Kumar2014b,Caprini2022,Farhadi2018,VanZuiden2016,lee_2017}.
As demonstrated, inertia plays a pivotal role in shaping emergent dynamics and can even impact steady-state properties within such systems. Recent research indicates the mitigation of motility-induced phase separation~\cite{Fily2012,Redner2013,Caporusso2020,omar23} and instability in active nematics~\cite{chatterjee2021inertia} with the introduction of inertia.

Moreover, consideration of confinement is often crucial for active matter, given that numerous biological processes occur within confined spaces, such as chromosomes within cell nuclei or cytoplasm within cells. Active particles exhibit intriguing properties when confined, such as aggregation at the confinement boundary, which differs from their passive counterparts~\cite{Caraglio_22,Nakul_23,Malakar_20,Dan_20,Hennes_14,Woillez_20,Basu_2020,Smith_22,Buttinoni_2022,Santra_21,Goswami_19,Takatori_16,Gutierrez_20,Muhsin_22,Nguyen_2022,Caprini_21,arredondo2023inertia,Frydel_24,obreque2024dynamics}.
Nevertheless, despite the tremendous progress in active matter research, the precise influence of inertia and confinement on the characteristics of active particles remains to be completely understood.

In this study, we analyze the motion of inertial ABPs (iABPs) confined within a harmonic trap. Through a Laplace transform of the governing Fokker-Planck equation, we demonstrate the precise time evolution of all moments of any desired dynamical variable in arbitrary $d$ dimensions—a primary accomplishment of this work. This method, initially devised for investigating worm-like chains in 1952~\cite{herman52}, has been recently adapted to study ABP dynamics~\cite{Chaudhuri_2021,Shee_20,Shee_pre_22,Shee_22,Patel_2023}. We provide the exact expressions for the time evolution of various position and velocity moments, validating them against direct numerical simulations. Both position and velocity moments converge to a steady state in the long time limit. 
Remarkably, key asymptotic physical quantities like steady-state kinetic temperature and an estimate of effective diffusion coefficient underscore the critical influence of both inertia and trap strength. Such properties have no equilibrium counterpart. 
Our calculations in trap suggest a possible inertia-dependence of diffusivity in a system of interacting iABPs, as has been observed in recent numerical simulations~\cite{khali2023does}.

In sharp contrast to equilibrium, the velocity and position distribution of iABPs in harmonic trap shows strong non-Gaussian departures that we capture exactly by deriving excess kurtosis in velocity and position. Remarkably, unlike passive particles, all the steady-state velocity and position moments of iABPs depend on inertia. We obtain `phase diagrams' showing departures from Gaussianity with the change in the three control parameters: active speed, trapping strength, and inertia. Note that, here, by transition, we denote dynamic crossovers between active and passive behaviors, as this single-particle system does not have any real phase transition.

The velocity kurtosis shows positive and negative departures associated with distributions of velocity being heavy-tailed-unimodal or having an inverted wine bottle or Mexican hat shape, respectively.
With activity, the velocity distribution becomes more and more non-Gaussian, where heavy-tailed distributions are preferred by lower inertia and higher trap strength. Larger inertia and shallower traps support the inverted Mexican hat-shape distributions. A clear re-entrance with increasing inertia is observed over a wide range of trap strengths.    

A qualitatively different `phase diagram' is obtained in position, using its excess kurtosis. It shows two `phases', one with negative values associated with an inverted Mexican hat-shaped distribution corresponding to particles climbing the trapping potential and the other with vanishing values corresponding to an equilibrium-like  Gaussian distribution of particle positions peaked at the center. This phase diagram shows re-entrance with increasing trap strength and a monotonic vanishing of excess kurtosis with increasing inertia.

The rest of the paper is organized as follows: Section \ref{sec_model} presents the model using Langevin dynamics and the Fokker-Planck equation-based method to compute all the dynamical moments in arbitrary dimensions. Section\,\ref{sec:firstorder} delves into the calculations of the first moments. Section\,\ref{sec:secondorder} presents the calculations of second moments and discusses quantities like kinetic temperature and diffusivity. Section\,\ref{sec:fourthorder} presents the calculation of fourth moments and discusses `phase diagrams' using kurtosis in velocity and displacement. 
Finally, Section\,\ref{sec_conclusion} presents an outlook summarizing our findings and discussing possible experiments to verify our predictions. 

\section{Model and Calculation of Moments}
\label{sec_model}
The underdamped dynamics of ABPs in the presence of a harmonic trap of strength $k$ in $d$ dimension is described by its position $\rv' = (r'_1,r'_2,...,r'_d)$, velocity $\vv' = (v'_1,v'_2,...,v'_d)$ and active velocity $v_a$ in the heading direction $\uv = (u_1,u_2,...,u_d)$ evolving with time $t'$.
Apart from an active speed $v_a$, the translational and rotational diffusivity $D$ and $D_r$ control the motion of ABPs.  
We use the unit of time $\t_r = 1/D_r$ and length $l = \sqrt{D/D_r}$ to express all other quantities. The speed and velocity, e.g., are expressed in units of $\bar{v} = \sqrt{DD_r}$. Using dimensionless time $t = t'/\t_r$, position $\rv = \rv'/l$, and velocity $\vv = \vv'/\bar{v}$, we express the Langevin equation in $d$ dimensions in the following Ito form~\cite{Ito1975,VandenBerg1985,Mijatovic2020} 
\bea
\label{eq:rlang}{\rm d} r_i &&= v_i \, {\rm d}t\\
\label{eq:vlang}M {\rm d}v_i &&= -( v_i -\Pe\, u_i + \beta r_i)\, {\rm d}t +\sqrt{2} \,{\rm d}B_i (t) \\
\label{eq:ulang}{\rm d}u_i &&= (\delta_{ij} - u_i u_j) \sqrt{2} \,{\rm d}B_j (t) - (d-1)u_i\, {\rm d}t.
\eea
In the above we used the following dimensionless parameters: trap strength $\beta = \mu k/ D_r$, inertial relaxation time $M = \t\, D_r$ where $\t = \mu\, m$ with mass $m$ and mobility $\mu$, and P{\'e}clet number $\Pe= v_a/\sqrt{DD_r}$.
The Gaussian white noise in translation and rotation obey $\la {\rm d}B_i \ra = 0$ and $\la {\rm d} B_i {\rm d} B_j \ra = \d_{ij} {\rm d t}$.
The first term on the right-hand side in equation~(\ref{eq:ulang}) projects the Gaussian noise on the $(d-1)$-dimensional hypersurface, and the second term ensures $\uv^2 = 1$ at all times. Alternatively, one can express this equation in the Stratonovich form $du_i = (\delta_{ij} - u_i u_j) \circ \sqrt{2} \,{\rm d}B_j (t)$.
The Langevin equation can be integrated numerically using the Euler-Maruyama scheme.

Noting that the position $\rv$ evolves deterministically via a velocity $\vv$, and the velocity, in turn, evolves via active and passive drift and diffusion, while the heading direction $\uv$ undergoes orientational diffusion on a $(d-1)$-dimensional hypersurface, we can write the corresponding Fokker-Planck equation for the probability distribution $P(\rv,\vv,\uv,t)$
\bea \fl
\label{eq:fp} \p_t P = - \nabla \cdot (\vv P) -(1/M) \nabla_v \cdot((\Pe\,\uv -\vv -\beta \, \rv)P) +(1/M^2) \nabla_v^2 P + \nabla_u^2P
\eea
where $\nabla$ and $\nabla_v$ are the gradient operators defined in arbitrary $d$-dimensions corresponding to position and velocity variables respectively and, $\nabla_u^2$ is a  spherical Laplacian defined over ($d-1$) dimensional hypersurface.
This operator can be expressed in the $d$-dimensional Cartesian coordinates $\textbf{y}$ as, $\nabla_u^2 = y^2 \sum_{i=1}^d \p_{y_i}^2 - [y^2\p_y^2 +(d-1)y\p_y]$.

Under Laplace transform of time, $\tilde{P} (\rv, \vv, \uv, s) = \int_{0}^{\infty} dt\, e^{-st}\,P(\rv, \vv, \uv, t)$, the Fokker-Planck equation takes the following form,
\bea \fl \label{eq:laplace}
-P(\rv, \vv, \uv, 0) + s \tilde{P}(\rv, \vv, \uv, s) &&= - \nabla \cdot (\vv \tilde{P}) - (1/M)\nabla_v \cdot ((\Pe\, \uv -\vv - \beta \, \rv)\tilde{P}) 
 \nn\\  &&+(1/M^2) \nabla_v^2 \tilde{P} + \nabla_u^2 \tilde{P}\,,
\eea
where $P(\rv, \vv, \uv, 0)$ sets the initial condition.
Defining the mean of any arbitrary observable $\psi$ as $\la \psi \ra_s = \int d\rv d\vv d \uv \, \psi(\rv, \vv, \uv)\, \tilde{P}(\rv, \vv, \uv, s)$ and using the above equation, we get 
\bea \fl \label{eq:observable}
 - \la \psi \ra_0 + s \la \psi \ra_s &&= \la \vv \cdot \nabla \psi\ra_s -(1/M) \la \vv \cdot \nabla_v \psi \ra_s + (\l/M) \la \uv \cdot \nabla_v \psi \ra_s - (\beta/M) \la \rv \cdot \nabla_v \psi \ra_s 
 \nn\\ 
&&+(1/M^2) \la \nabla_v^2 \psi \ra_s + \la \nabla_u^2 \psi \ra_s,
\eea
where $\la \psi \ra_ 0 = \int d\rv d\vv d\uv\, \psi(\rv, \vv, \uv) \, P(\rv, \vv, \uv, 0)$ and the initial condition $P(\rv, \vv, \uv, 0)= \d(\rv-\rv_0) \d(\vv - \vv_0)\d(\uv-\uv_0)$. 
Equation\,(\ref{eq:observable}) can be used to get the exact expression for the time evolution of any arbitrary observable by taking an inverse Laplace transform.
In the following sections, we present several examples of such calculations.
We denote the steady-state values $\lim_{t \rightarrow \infty} \la \psi \ra (t) = \lim_{s\rightarrow 0}s \la \psi \ra_s  \equiv \la \psi \ra_{\rm st}$. 

\begin{figure}
    \centering
    \includegraphics[scale = 0.14]{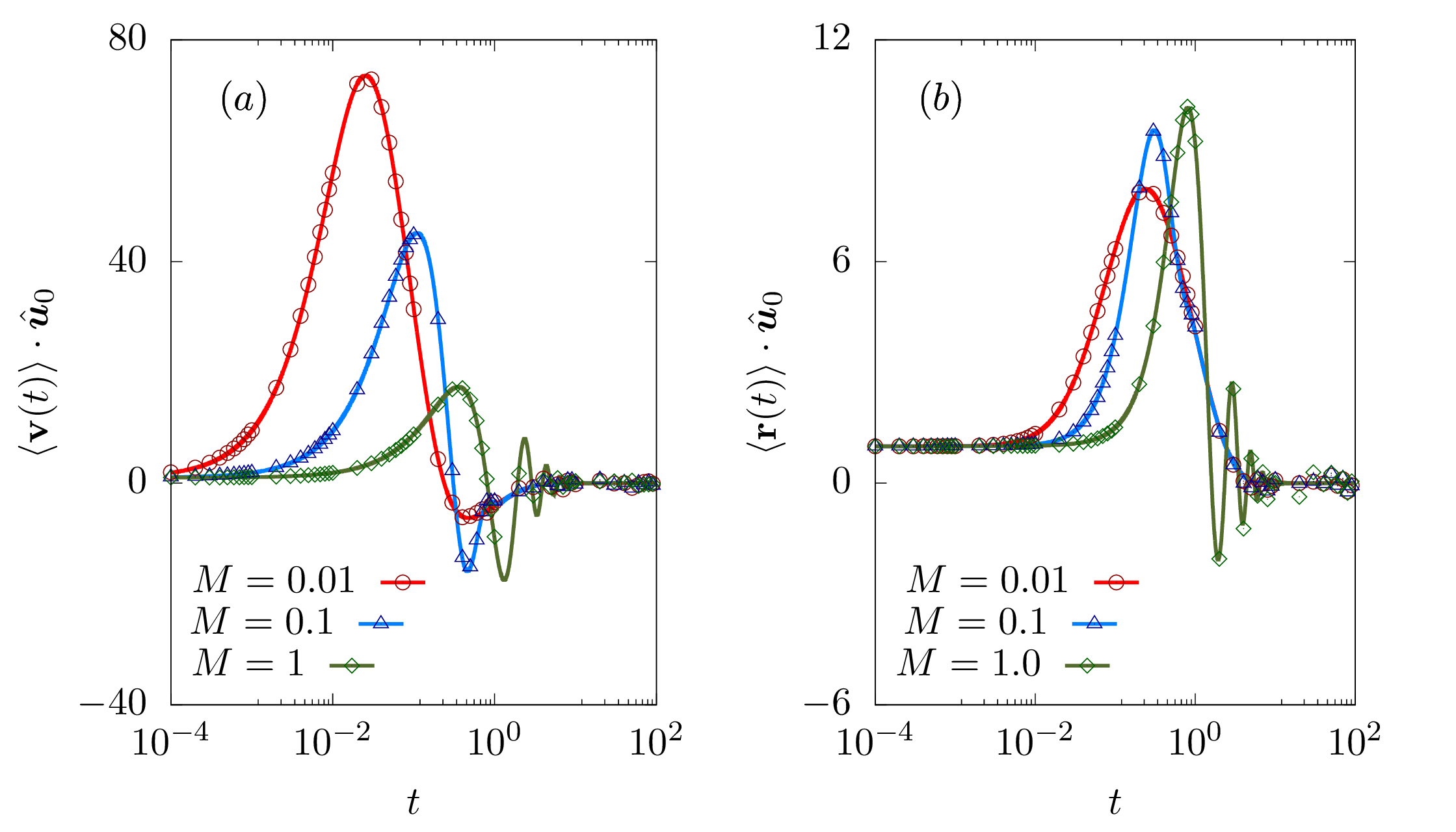}
    \caption{Time evolution of first moments $\la \vv(t) \ra $ ($a$) and $\la \rv(t)  \ra$ ($b$) projected along initial orientation $\uv_0$ in $d = 2$ dimensions.
    The solid lines are obtained from equations\,(\ref{eq:vt}) and (\ref{eq:rt}), respectively. The points denote simulation results.
    The plots correspond to $\Pe = 100$, $\beta = 10$, and initial conditions $\rv_0 = (1,0)$, $\vv_0 = (1,0)$, and $\uv_0 = (1,0)$. }
    \label{fig:firstmoments}
\end{figure}


\section{First moments}
\label{sec:firstorder}
We begin by utilizing equation\,(\ref{eq:observable}) to obtain the time evolution of several first-order moments, such as the mean velocity, position, and heading direction. In the presence of a harmonic trap, the dynamics of velocity and position both evolve to reach a steady state. 
\subsection{Mean velocity, position and heading direction}
Utilizing $\psi = \vv$ in equation\,(\ref{eq:observable}), we get the mean velocity in the Laplace space $\la \vv \ra_s = [\vv_0 + (\l/M) \la \uv \ra_s - (\beta /M) \la \rv \ra_s ]/(s+1/M)$ in the terms of $\la \rv \ra_s$ and $\la \uv \ra_s$.
Again, using $\psi = \rv$ and $\psi = \uv$, equation\,(\ref{eq:observable}) gives $\la \rv \ra_s = [ \la \vv \ra_s + \rv_0 ]/s$ and $\la \uv \ra_s = \uv_0/(s+d-1)$, respectively.
The solution of these coupled equations leads to
\bea
\label{eq:vs} \la \vv \ra_s = \frac{(sM \vv_0 - \beta \,\rv_0)}{Ms^2 +s + \beta} + \frac{s \Pe\, \uv_0 }{(s+d-1)(Ms^2+s+\beta)},\\
\label{eq:rs} \la \rv \ra_s =   \frac{M \vv_0 + (1 + sM) \rv_0}{Ms^2 +s + \beta} + \frac{\Pe\, \uv_0}{(s+d-1)(Ms^2+s+ \beta)}.
\eea
The inverse Laplace transform of equation\,(\ref{eq:vs}) and $\la \uv \ra_s$ gives the full-time evolution which can be expressed as
\bea
\label{eq:vt}\fl \la \vv \ra (t) = - \frac{(d-1)\Pe \, \uv_0}{{\cal C}_1} e^{-(d-1)t} + \frac{1}{2} \left( \vv_0 + \frac{{(d-1)\rm Pe} \,\uv_0}{{\cal C}_1} \right) \left[ e^{- \alpha_{+}t/2} + e^{-\alpha_{-}t/2} \right] \nn\\
\fl  + \frac{1}{2 i \omega {\cal C}_1} \left\{ (d-1)[\Pe \, \uv_0  + ((d-1)M-1)\vv_0] + \beta [2(d-1)((d-1)M-1)\rv_0 \right. \nn\\
\fl \left. - 2\Pe \, \uv_0 + \vv_0]  + 2 \beta^2 \rv_0 \right\}   \left[ e^{- \alpha_{+}t/2} - e^{-\alpha_{-}t/2}  \right],
\eea
where
\bea
\omega = \sqrt{4 \beta M - 1},\nn\\
\alpha_{\pm} = \frac{1}{M}[1\pm i \omega],\nn\\
{\cal C}_1 = (d-1)^2M-(d-1)+\beta,
\eea
and $\la \uv \ra (t) = \uv_0 e^{-(d-1)t}$.
The mean velocity vanishes in the steady-state $\la \vv \ra_{\rm st} = 0$. Before vanishing, it shows an oscillatory decay with frequency $\w/M$ if $\be M >1/4$, the imaginary part of $\alpha_{\pm}$.   The case of $4 \beta M = 1$ corresponds to critical damping in oscillations.  
In the limit of vanishing trap strength, equation\,(\ref{eq:vt}) reduces to the known expression for free iABPs~\cite{Patel_2023}.

Likewise, the inverse Laplace transform of equation\,(\ref{eq:rs}) gives the full-time evolution of mean position as
 \bea
 \label{eq:rt}
 \fl \la \rv \ra (t) = \frac{\Pe\,\uv_0}{{\cal C}_1} e^{-(d-1)t} + \frac{1}{2 i \omega} \left( \rv_0 + 2M \vv_0 + \frac{[2M(d-1)-1] \Pe\, \uv_0 }{{\cal C}_1} \right)  \left[ e^{-\alpha_{-}t/2} \right. \nn\\
 \fl \left. - e^{-\alpha_{+}t/2}  \right]+\frac{1}{2} \left( \rv_0 - \frac{\Pe\, \uv_0}{{\cal C}_1} \right) \left[ e^{-\alpha_{-}t/2} + e^{-\alpha_{+}t/2} \right]. 
 \eea
 In the steady state, the mean position also vanishes $\la \rv \ra_{\rm st} = 0$ and shows an oscillatory evolution towards the steady state with frequency $\w/M$ if $\be M > 1/4$, as for mean velocity.
 In the limit of vanishing inertia, equation\,(\ref{eq:rt}) restores the previously known result of overdamped trapped ABPs~\cite{Chaudhuri_2021}, and in the vanishing trap strength, it restores the result of free iABPs~\cite{Patel_2023}.
 
{In figure~\ref{fig:firstmoments}, we show projections of mean velocity and position along the initial orientation evolving towards the steady state. 
Two distinct kinds of evolutions are observed. For $4 \beta M > 1$, $\alpha_\pm$ has an imaginary part, which sets oscillation with $\omega/M$ frequency and a real part which governs the decay in time with a time constant $2M$ for both $\la \rv \ra (t)$ and $\la \vv \ra (t)$, as shown for $M = 0.1, 1$ in figure~\ref{fig:firstmoments} in two dimensions (2d). However, at $4 \beta M < 1$, $\alpha_\pm $ is real, and the solution of $\la \rv \ra (t)$ and $\la \vv \ra (t)$ decays to zero asymptotically in a non-monotonic manner that lacks oscillation. This is shown for $M = 0.01$ in figure~\ref{fig:firstmoments}.
The moments $\la \rv(t) \ra \cdot \uv_0$ and $\la \vv(t) \ra \cdot \uv_0$ starts from $\uv_0 \cdot \rv_0$ and $\uv_0 \cdot \vv_0$, respectively at the short time.
From equations~(\ref{eq:vs}) and (\ref{eq:rs}), it is easy to see that, in the presence of the trap, the steady-state values vanish, $\la \vv \ra_{\rm st} = \lim_{s \to 0+} s \la \vv \ra_s =0$ and $\la \rv \ra_{\rm st} = \lim_{s \to 0+} s \la \rv \ra_s =0$. 
}

\section{Second Moments: Mean-Squared Displacement and Kinetic Temperature}
\label{sec:secondorder}
In this section, we explore the time evolution of second moments, including the projection of position and velocity towards instantaneous heading direction, mean squared displacement~(MSD), and mean squared velocity~(MSV). Note that the projections involve cross-correlation of position and velocity with orientation.

\subsection{Position and velocity component in heading direction}

The simplest second moments are inner products of velocity and displacement variables with instantaneous orientation, in other words, their components along the heading direction. We use $\psi = v_{\parallel} = \uv \cdot \vv $ in equation\,(\ref{eq:observable}) to get $[s+1/M+(d-1)] \la v_{\parallel} \ra_s = \uv_0 \cdot \vv_0 + \l/(sM) - (\beta/M) \la \uv \cdot \rv \ra_s , $ in terms of $\la \uv \cdot \rv \ra_s$. 
Again using $\psi = r_{\parallel} = \uv \cdot \rv $ in equation\,(\ref{eq:observable}), we get $[s+(d-1)] \la r_{\parallel} \ra_s = \uv_0 \cdot \rv_0 +\la v_{\parallel} \ra_s$.
 The solution of these coupled equations gives $\la v_{\parallel} \ra_s$ and $\la r_{\parallel} \ra_s$ as 
 \bea
\label{eq:uvs} \la v_{\parallel} \ra_s = \frac{[s+(d-1)] (\Pe + s M \uv_0 \cdot \vv_0) - s \beta \uv_0 \cdot \rv_0 }{s[\beta + (s+d-1) \{M(s+d-1) +1 \}]}, \\
\label{eq:urs} \la r_{\parallel} \ra_s = \frac{\Pe+ s \left[ \{M(s+d-1) +1 \} \uv_0 \cdot \rv_0 + M \uv_0 \cdot \vv_0 \right] }{s[\beta + (s+d-1) \{M(s+d-1) +1 \}]}.
 \eea
 The inverse Laplace transform  of the above equations give
\bea
\label{eq:uvt}
\fl \la v_{\parallel} \ra (t) = \frac{(d-1)\Pe}{{\cal C}_2} + \frac{1}{2 i \omega} \left[ \frac{(d-1+2\beta) \Pe}{{\cal C}_2} - 2\beta \uv_0 \cdot \rv_0  - \uv_0 \cdot \vv_0 \right] \left( e^{-\{\alpha_{-} +2(d-1)\}t/2} \right. \nn\\
\fl \left. - e^{-\{\alpha_{+} +2(d-1)\}t/2 }  \right)  - \frac{1}{2} \left[ \frac{(d-1)\Pe}{{\cal C}_2}  - \uv_0 \cdot \vv_0 \right] \left( e^{-\{\alpha_{-} +2(d-1)\}t/2 } + e^{-\{\alpha_{+} +2(d-1)\}t/2}  \right) ,
\eea
and
\bea
\label{eq:urt}
\fl \la r_{\parallel} \ra (t) = \frac{\Pe}{{\cal C}_2} + \frac{1}{2 i \omega} \left[ \frac{(1+2M(d-1)) \Pe}{{\cal C}_2} - \uv_0 \cdot \rv_0 - 2M \uv_0 \cdot \vv_0 \right] \left( e^{-\{\alpha_{+} +2(d-1)\}t/2} \right. \nn\\
\fl \left. - e^{-\{\alpha_{-} +2(d-1)\}t/2}   \right) - \frac{1}{2} \left[ \frac{\Pe}{{\cal C}_2} - \uv_0 \cdot \rv_0 \right] \left( e^{-\{\alpha_{-} +2(d-1)\}t/2 } + e^{-\{\alpha_{+} +2(d-1)\}t/2}  \right),
\eea
with
\bea
{\cal C}_2 = (d-1)^2M+(d-1)+\beta.
\eea
In the asymptotic limit, the two projections $\la v_{\parallel} \ra$ and $\la r_{\parallel} \ra$ approach the steady state values $\Pe/[(d-1)M+1+\beta/(d-1)]$ and $\Pe/[(d-1)^2M+d-1+\beta]$ respectively.
In the vanishing trap strength limit, $\la v_{\parallel} \ra_{\rm st}$ takes the previously known form of free iABPs~\cite{Patel_2023}. In the other limit of $\beta \to \infty$, both projections $\la r_{\parallel} \ra_{\rm st}$ and $\la v_{\parallel} \ra_{\rm st}$ vanishes.

 \begin{figure}
    \centering    \includegraphics[scale=0.60]{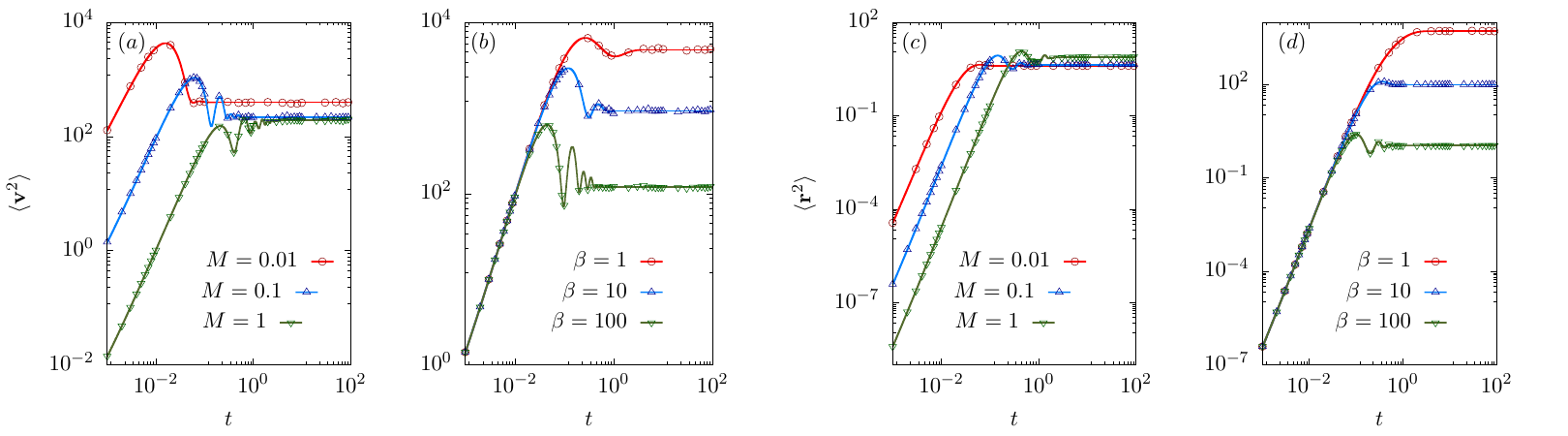}
    \caption{Time evolution of second moments $\la \vv^2 \ra(t)$ and $\la \rv^2 \ra (t)$ in $d$ = 2.
    We show the evolution of $\la \vv^2 \ra$ and $\la \rv^2 \ra$ for different $M$ values at $\beta = 50$ ($a$), ($c$) and for different $\beta$ values at $M = 0.1$ ($b$), ($d$). 
    The plots use $\Pe = 100$, $\rv_0 = (0,0)$, $\vv_0 = (0,0)$ and $\uv_0 = (1,0)$. The lines correspond to equation\,(\ref{eq:v2t}) in ($a$), ($b$) and equation\,(\ref{eq:r2t}) in ($c$), ($d$). The points denote simulation results.  }
    \label{fig:secondorder}
\end{figure}
\subsection{Mean Squared Velocity}
We set $\psi =  \vv^2 $ in equation\,(\ref{eq:observable}) to find the second moment of velocity.
We use the following relations: $\vv \cdot \nabla \psi = 0$, $\la \vv \cdot \nabla_v \psi \ra_s = 2\,\la \vv^2 \ra_s$, $\la  \uv \cdot \nabla_v \psi \ra_s = 2\,\la v_{\parallel} \ra_s$, $ \la \rv \cdot \nabla_v \psi \ra_s = 2 \,\la \vv \cdot \rv \ra_s$, $\la \nabla_v^2 \psi \ra_s = 2d \la 1 \ra_s = 2d/s$ ~\footnote{As, $\la 1 \ra_s = \int d\rv d \vv d\uv \tilde{P} = \int d\rv d\vv d\uv \int_{0}^{\infty} dt e^{-st}P = \int_{0}^{\infty}dte^{-st}\{\int d\rv d\vv d\uv P \} = \int_{0}^{\infty}dt e^{-st} = 1/s.$} and $\nabla_u^2 \psi = 0$, to get 
\bea
\label{eq:v2s}
\fl (s+2/M) \la \vv^2 \ra_s = \vv_0^2 + (2\,\l/M)\la v_{\parallel} \ra_s - (2\beta/M)\la\vv \cdot \rv \ra_s + 2d/(sM^2),
\eea
where $ \la v_{\parallel} \ra_s$ is already known from equation\,(\ref{eq:uvs}). We further calculate  $ \la \vv \cdot \rv \ra_s$, by inserting $\psi = \vv \cdot \rv$ in equation\,(\ref{eq:observable}) to get
\bea
 \label{eq:vrs}(s+1/M) \la \vv \cdot \rv \ra_s = \vv_0 \cdot \rv_0 +\la \vv^2 \ra_s + (\l/M) \la r_{\parallel} \ra_s - (\beta/M) \la \rv^2 \ra_s ,
\eea
where $\la r_{\parallel} \ra_s$ is given by equation\,(\ref{eq:urs}). Completing the calculation further requires the second moment of position $\la \rv^2 \ra_s$. Using $\psi = \rv^2$ in equation \,(\ref{eq:observable}) we get
\bea
\label{eq:r2s}s \la \rv^2 \ra_s = \rv_0^2+ 2 \la \vv \cdot \rv \ra_s.
\eea
Equations\,(\ref{eq:v2s}) to (\ref{eq:r2s}) give the required moments in the Laplace space. For the choice of initial condition $\vv_0 = \bf{0}$ and $\rv_0 = \bf{0}$, the expression simplifies significantly and gives the MSV
\bea
\label{eq:v2sf}\fl \la \vv^2 \ra_s = \frac{2}{{\cal B}_0} \left( s(s+d-1)(sM+1)[d(d+1)M+M\Pe^2+dM(s-1)]  + 2d \beta^2 \right. \nn\\
\fl \left. + \beta \left[ 2(d-1)\{d((d-1)M+1) + M \Pe^2 \} +sd\{3+4M(d-1)\}  +sM\Pe^2 + 3dM s^2  \right] \right),\nn\\
\eea
and cross-correlation
\bea
\label{eq:vrsf}\fl \la \vv \cdot \rv \ra_s = \frac{sM}{{\cal B}_0} \left( \Pe^2 [(3s-2)M+2 ] + 2d [s+\beta -1 + M((s-1)^2 +\Pe^2) ] \right. \nn\\ 
\left. \fl + d^2 [4M(s-1)+2 ] + 2d^3M     \right),
\eea
where ${\cal B}_0 = sM(sM+1)\{ s(sM+2)+4\beta \} \{ (s+d-1)[(s+d-1)M+1] + \beta \}$. It is easy to check that the steady state cross-correlation $\la \vv \cdot \rv \ra_{st} = \lim_{s \to 0+} s \la \vv \cdot \rv \ra_s = 0$. 

The inverse Laplace transform  of equation\,(\ref{eq:v2sf}) gives the full-time evolution of MSV,
\bea
\label{eq:v2t}\fl \la \vv^2 \ra (t) =  \frac{d}{M} +\frac{(d-1) \Pe^2}{{\cal C}_2}  + \frac{2\beta }{\omega^2} \left( -2d + \frac{[1-2(d-1)M]\Pe^2}{{\cal C}_1} \right) e^{-t/M} \nn\\
\fl -\frac{(d-1)\Pe^2}{i \omega {\cal C}_1 {\cal C}_2} \left([2\beta + (d-1)M\alpha_{-} ] e^{-\{\alpha_{-} +2(d-1)\}t/2}  - [2\beta + (d-1)M\alpha_{+}  ]  e^{-\{\alpha_{+} +2(d-1)\}t/2}\right) \nn\\
\fl  + \frac{1}{2 \omega^2{\cal C}_1} \left(  d  \alpha_{-} \left[ {\cal C}_1 +M \Pe^2\right] - \Pe^2 \{ M \alpha_- + 2 \beta \}  \right) e^{-\alpha_{-}t} \nn\\
\fl + \frac{1}{2 \omega^2{\cal C}_1} \left(  d  \alpha_{+} \left[ {\cal C}_1 +M \Pe^2\right] - \Pe^2 \{ M \alpha_+ + 2 \beta \}  \right)e^{-\alpha_{+} t} .
\eea
As shown in the figure~\ref{fig:secondorder}($a$,$b$), the plot of the above equation agrees with the numerical simulation results of MSV in 2d. The evolution in harmonic trap shows oscillations before saturating to steady state values determined by $\Pe$, $M$, and $\be$.  In the limit of vanishing trap strength, MSV takes the form of free iABPs~\cite{Patel_2023}. The short-time evolution can be further analyzed using an expansion of $\la \vv^2 \ra (t)$ around $t = 0$ 
\bea
\fl \la \vv^2 \ra (t) = \frac{2d}{M^2}t + \frac{(M\Pe^2 - 2d)}{M^3}t^2 - \frac{[2d(\beta M -2) +M\,\Pe^2\{ 3+(d-1)M\}]}{3M^4} t^3 + {\cal O}(t^4) \nn\\
\eea
It takes time, determined by inertia and trap strength, before the influence of activity, and subsequently, the harmonic trap shows up in the evolution of $\vv^2(t)$. MSV shows a diffusive scaling similar to free iABPs in the shortest time~\cite{Patel_2023}. 
The expansion shows a diffusive $\la \vv^2 \ra \sim t$ to ballistic $\la \vv^2 \ra \sim t^2$ crossover at $t_I  = 2dM/(M\Pe^2-2d)$, which depends on $M$ but is independent of $\beta$. At high $\Pe$, it shows non-monotonicity at a later time and decreases as $\la \vv^2 \ra \sim\, -t^3$ beyond  $t_{II} = (3M(\Pe^2 M -2d))/[2d(\beta M -2) +M\,\Pe^2\{ 3+(d-1)M\}]$~(figure\,\ref{fig:secondorder}($a$)\,). This crossover time depends on the trap strength $\beta$, which also causes subsequent oscillations before reaching a steady state. 
The asymptotic steady-state value of MSV is given by 
\bea \label{eq_v2st}
    \la \vv^2 \ra_{\rm st} = \frac{d}{M} +\frac{(d-1) \Pe^2}{(d-1)^2M+ (d-1)+ \beta }, 
\eea
which decreases with both $M$ and $\beta$. 
In the limit of $\beta \rightarrow 0$, $\la \vv^2 \ra_{\rm st}$ reduces to the previously known form of free iABPs~\cite{Patel_2023}.

In figure~\ref{fig:ss_second}(a) and (b), parameter dependence of steady-state MSV, $\la \vv^2 \ra_{\rm st}$, are shown. They start from maximum value $d/M+\Pe^2/[(d-1)M+1]$ at small $M$ and $\beta$ to decrease with increasing inertia and trap stiffness. Since $\la \vv \ra_{\rm st} = 0$, the steady-state velocity fluctuation is also given by the MSV, $\la \delta \vv^2 \ra_{\rm st} = \la \vv^2 \ra_{\rm st}$.

\subsection{Mean Squared Displacement}
\begin{figure}
    \centering
    \includegraphics[scale = 0.55]{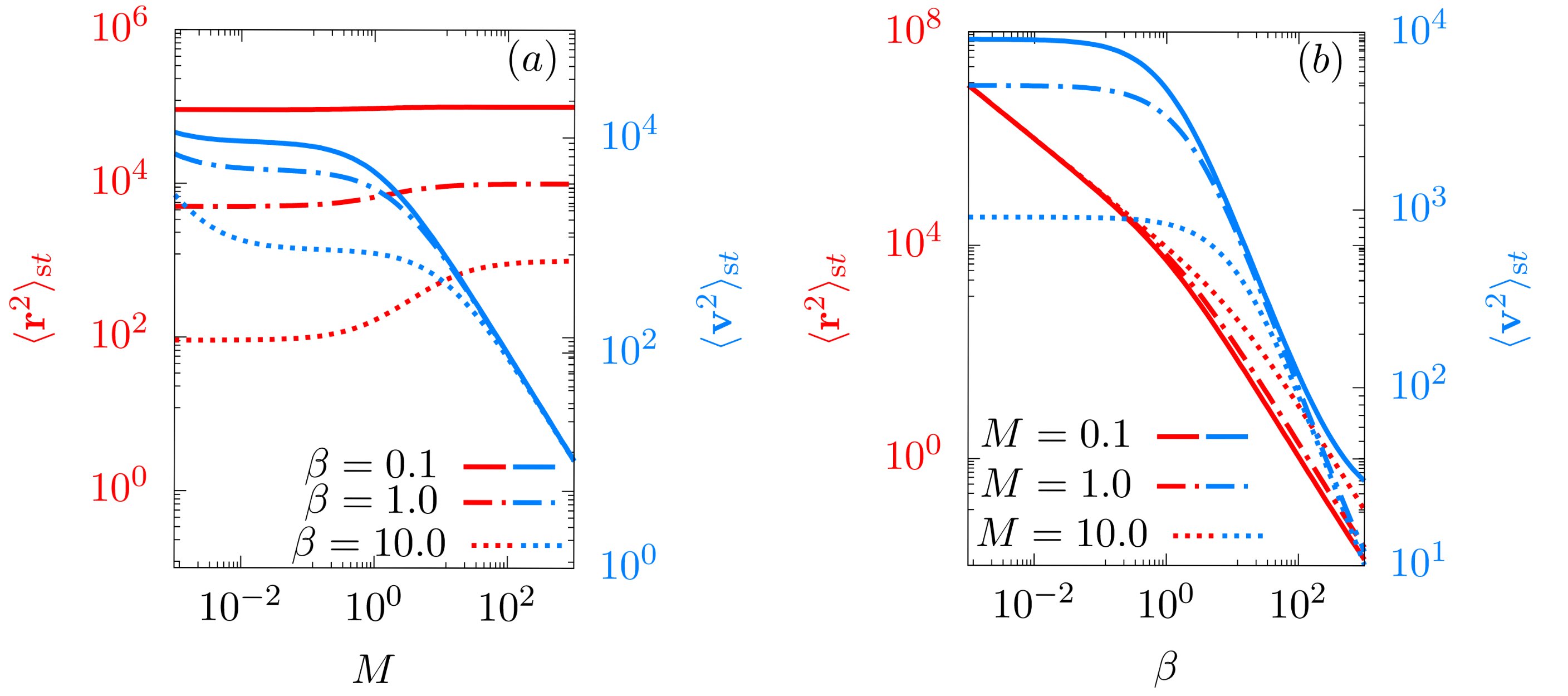}
    \caption{The steady-state of second moments $\la \vv^2 \ra_{\rm st}$~(blue), $\la \rv^2 \ra_{\rm st}$~(red) as a function of $M$ ($a$) and $\beta$ ($b$) in $2$d. The lines are plots of equations\,(\ref{eq_v2st}) and (\ref{eq:r2st}) at $\Pe = 100$. 
    }
    \label{fig:ss_second}
\end{figure}

Using equations\,(\ref{eq:v2s}) to (\ref{eq:r2s}) we obtain the MSD in Laplace space. With the initial condition $\vv_0 = \bf{0}$ and $\rv_0 = \bf{0}$, the expression for MSD simplifies to
\bea
\fl \la \rv^2 \ra_s = \frac{M}{{\cal B}_0} \left(4d\{ s-1+\beta+M[(s-1)^2 +\Pe^2] \} + d^2 [8M(s-1) +4]  + 4 d^3M \right. \nn\\
\fl \left.+ 2 \Pe^2 [M(3s-2)+2] \right)
\eea
where ${\cal B}_0$ has the same $s$-dependent form as used before to describe $\la \vv^2 \ra_s$.
The inverse Laplace transform  of this relation reads
\bea
\label{eq:r2t}\fl \la \rv^2 \ra (t) = \frac{d}{\beta} + \frac{\Pe^2(1+(d-1)M)}{\beta{\cal C}_2} 
+ \frac{2M }{\omega^2} \left( -2d + \frac{(1-2M(d-1))\Pe^2 }{{\cal C}_1} \right) e^{-t/M} \nn\\
\fl - \frac{\left[\alpha_{+}+2(d-1) \,\right] M \Pe^2}{i \omega {\cal C}_1{\cal C}_2} e^{-\{\alpha_{-}+2(d-1)\}t/2}  + \frac{\left[\alpha_{-}+2(d-1) \,\right] M\Pe^2}{i \omega {\cal C}_1{\cal C}_2} e^{-\{\alpha_{+}+2(d-1)\}t/2} \nn\\
\fl +\frac{1}{2 i \beta \omega^3 {\cal C}_1} \left( i \omega d M \alpha_{-} \left[ {\cal C}_1 + M \Pe^2 \right]   + \Pe^2 \left[ M \alpha_-(1+M-2 \beta M) -2 \beta M (1 + 2M) \right] \right) e^{-\alpha_{+}t}\nn\\
\fl +\frac{1}{2 i\beta \omega^3 {\cal C}_1} \left( i \omega d M \alpha_{+} \left[ {\cal C}_1 + M \Pe^2 \right]   - \Pe^2 \left[ M \alpha_+ (1+M - 2 \beta M)-2 \beta M(1+2M) \right] \right) e^{-\alpha_{-}t}.\nn\\
\eea
The above equation captures the evolution of MSD as seen in numerical simulation in $2$d, shown in figure \ref{fig:secondorder}($c$,$d$). The expression of MSD reduces to that of the overdamped case in the limit of vanishing inertia~\cite{Chaudhuri_2021} while it restores the result of free iABPs in the limit of vanishing trap strength~\cite{Patel_2023}. To analyze the crossovers at a short time, we expand $\la \rv^2 \ra$ around $t = 0$ 
\bea
\fl \la \rv^2 \ra (t) = \frac{2d}{3M^2}t^3 + \frac{(M\Pe^2 - 2d)}{4M^3}t^4 - \frac{ \{ d(4 \beta M-7) + M \,\Pe^2 [5+2(d-1)M] \} } {30M^4}t^5 + {\cal O}(t^6). \nn\\
\eea
MSD shows $\sim t^3$ scaling at the shortest time scale similar to free iABPs~\cite{Patel_2023}. The influence of activity shows up in the form of a $\la \rv^2 \ra \sim t^3$ to $\la \rv^2 \ra \sim t^4$ crossover at time $t_{I} = 8dM/[3(M\Pe^2 - 2d)]$. A subsequent crossover towards saturation of MSD appears approximately 
at $t_{II} \approx 15M(2d - \Pe^2 M )/(2 \{ d(4 \beta M - 7) + M \,\Pe^2 [5+2(d-1)M] \} )$, due to the trap and as a result depends on the trap strength $\beta$. In the asymptotic limit, MSD reaches the steady state value 
\bea \label {eq:r2st}
\la \rv^2 \ra_{\rm st} = \frac{d}{\beta} + \frac{\Pe^2(1+(d-1)M)}{\beta[(d-1)^2M +(d-1) + \beta]} \,,
\eea
which depends on both $M$ and $\beta$. In the limit of vanishing $M$, the above expression for $\la \rv^2 \ra_{\rm st}$ reduces to the previously known form of MSD in overdamped trapped ABPs~\cite{Chaudhuri_2021}. The MSD increases with $M$ to saturate to $\lim_{M \to \infty}[\la \rv^2 \ra_{\rm st}] = (d/\be + \Pe^2/\be(d-1))$. In figure~\ref{fig:ss_second}(a) and (b), such dependence of steady-state MSD, $\la \rv^2 \ra_{\rm st}$, are shown. With $M$, it increases to saturate to $(d/\be + \Pe^2/\be(d-1))$. However, with $\be$, it monotonically decreases from the $1/\be$ divergence at $\be \to 0$ to vanishing as $1/\be^2$ in the large $\be$ limit.   

\subsection{Estimates of kinetic temperature, diffusivity, and violation of equilibrium fluctuation-dissipation relation}

The steady-state expressions for MSV and MSD, derived above, can be used to obtain asymptotic estimates of kinetic temperature and  diffusivity of iABP in traps. The fluctuation in velocity leads to the following estimate for   kinetic temperature,
\bea
\label{eq:kinetic_temperature}k_{\rm B} T_{\rm kin} = \frac{M \la \vv^2 \ra_{\rm st}}{d} = 1 + \frac{M(d-1)\Pe^2}{d[(d-1)^2M+ (d-1)+ \beta]}.
\eea
In contrast to trapped Brownian motion in equilibrium, this estimate depends explicitly on inertia $M$ and trap strength $\be$.

In the vanishing $\beta$ limit, the expression agrees with the earlier estimate for free iABPs~\cite{Patel_2023}. In the overdamped limit $M \rightarrow 0$, the dimensionless $k_B T_{\rm kin } = 1$, set by translational diffusivity. At small $M$, $k_B T_{\rm kin }$ increases linearly with $M$ to saturate to $1 + \Pe^2/[d(d-1)]$, independent of $\be$. However, at a given $M$, the dimensionless $k_B T_{\rm kin}$ decreases with $\beta$ to saturate to unity.

The fluctuation in position at equilibrium gives $\la \rv^2\ra_{eq}=d\, D/\be D_r$ with equilibrium fluctuation-dissipation relation (FDR)  $D=\mu k_B T$. In the absence of such FDR, the out-of-equilibrium steady-state fluctuation can not be associated with such a unique definition of temperature. However, extending this notion, one can obtain the following estimate for the effective diffusivity  
\bea \label{eq:deff}
D_{\rm eff} := \frac{\beta \la \rv^2 \ra_{\rm st}}{d} = 1 + \frac{\Pe^2}{d[(d-1)+\frac{\beta}{1+(d-1)M}]},
\eea
in the dimensionless form, replacing $D_{\rm eff}/D$ by $D_{\rm eff}$ and using unity for $l^{-2}=D_r/D$ as $l$ sets the unit of length in the first expression presented above. 
It is important to note that $D_{\rm eff}$ depends on both $M$ and $\beta$, where the $M$ dependence enters through the potential strength $\be$.  Remarkably, numerical simulations of interacting iABPs display inertia-dependent diffusivity~\cite{khali2023does}, a behavior qualitatively similar to the above relation.    

In the free particle limit of $\be \to 0$, the active part of diffusivity reduces to $\Pe^2/d(d-1)$, a result established before for free iABP~\cite{Patel_2023} and indistinguishable from overdamped ABPs. Moreover, in the vanishing $M$ limit, $D_{\rm eff} = 1 + \Pe^2/[d(\beta + d-1)]$. With increasing $M$, $D_{\rm eff}$ increases to saturate to the free iABP diffusivity $1+\Pe^2/[d(d-1)]$.

The departure from equilibrium FDR in the dimensionless form can be expressed as 
\bea
{\cal I} := D_{\rm eff} - k_{\rm B} T_{\rm kin} = \frac{\Pe^2}{d[(d-1)^2M +(d-1) + \beta ]},
\eea
where in the first relation, we replaced $\mu k_{\rm B} T_{\rm kin}/D$ by the dimensionless form of kinetic temperature $k_{\rm B} T_{\rm kin}$.  
The violation of equilibrium FDR is the maximum for vanishing $M$ and $\beta$. It decreases with increasing $M$ and $\beta$ to vanish as $M^{-1}$ and $\beta^{-1}$. In the vanishing $\beta$ limit, the above expression reduces to the known form for free iABPs~\cite{Patel_2023}.

\begin{figure}
    \centering
    \includegraphics[scale=0.60]{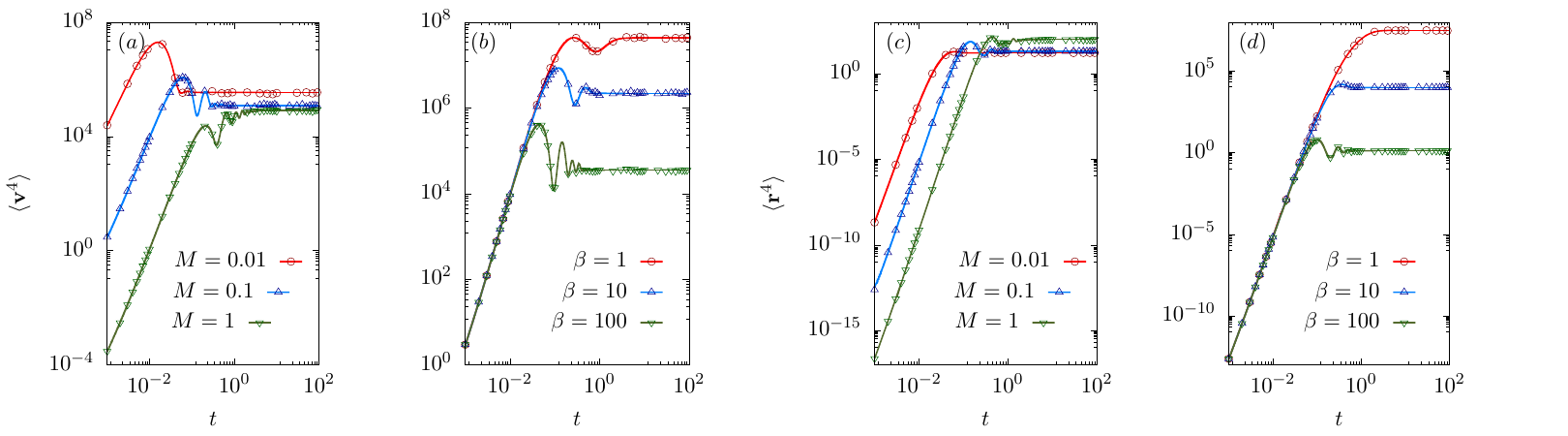}
    \caption{Time evolution of fourth moments $\la \vv^4 \ra(t)$ and $\la \rv^4 \ra (t)$ in $d$ = 2.
    The evolution of $\la \vv^4 \ra$ and $\la \rv^4 \ra$ for different $M$ values at $\beta = 50$ ($a$), ($c$) and for different $\beta$ values at $M = 0.1$ ($b$), ($d$). 
    The plots use $\Pe = 100$, $\rv_0 = (0,0)$, $\vv_0 = (0,0)$ and $\uv_0 = (1,0)$. The lines correspond to analytic estimates, and points denote simulation results.}
    \label{fig:fourthorder}
\end{figure}

\section{Fourth Moments}\label{sec:fourthorder}
In this section, we discuss the time evolution of fourth moments of dynamics, such as velocity and displacement, and the dependence of their steady-state values on control parameters.
A stochastic process can be distinguished from the Gaussian process using kurtosis, which involves the fourth moment of the corresponding random variable. We present such analyses for velocity and displacement with the help of excess kurtosis.

\subsection{Fourth Moment of Velocity and Displacement}
The calculation for fourth moments in Laplace space, $\la \rv^4 \ra_s$ and $ \la \vv^4 \ra_s$, follows the same steps shown before for other moments and utilizes equation\,(\ref{eq:observable}).
Despite being lengthy, these calculations are straightforward. We show the essential steps for the calculations in Laplace space in \ref{sec:fourthorder_appendix}. The inverse Laplace transform gives the full-time evolution. Instead of showing these long expressions for time-dependence, we plot them for $ \la \rv^4 \ra (t) $ and $ \la \vv^4 \ra (t)$ and compare them with numerical simulation results in $2$d in figure~\ref{fig:fourthorder}. The results are shown for a range of $M$ and $\beta$ values at $\Pe=100$. The simulation results show excellent agreement with analytic estimates. Apart from plotting the expressions, we present series expansions of them around $t=0$ to analyze their short-time behavior and also present the explicit steady-state expressions.
The series expansion of $\la \vv^4 \ra (t)$ around $t = 0$ in 2d (see \ref{sec:3dresults} for $d=3$ dimensions) using initial values $\rv_0 = \bf{0}$ and $\vv_0  = \bf{0}$ is given by
\bea
\label{eq:v42dser} \la \vv^4 \ra (t) &&= \frac{32}{M^4}t^2 + \frac{16(M\Pe^2 -4)}{M^5} t^3 \nn\\
&&+ \frac{[224+M\{-64 \beta -16(6+M)\Pe^2 + 3M \Pe^4 \}]}{3 M^6}t^4 +{\cal O}{(t^5)}.
\eea
The expansion shows an initial $t^2$ dependence for a short time, which crosses over to $t^3$ behavior at $t_I = 2M/(M \Pe^2 -4 )$ which depends on $M$ but independent of $\be$, a behavior similar to free iABPs~\cite{Patel_2023}. The impact of trapping is felt only at a later time $t_{II} = 48M(M \Pe^2- 4)/[224+M\{-64 \beta -16(6+M)\Pe^2 + 3M \Pe^4 \}]$ leading to a departure from the $t^3$ behavior.

\begin{figure}
    \centering
    \includegraphics[scale = 0.6]{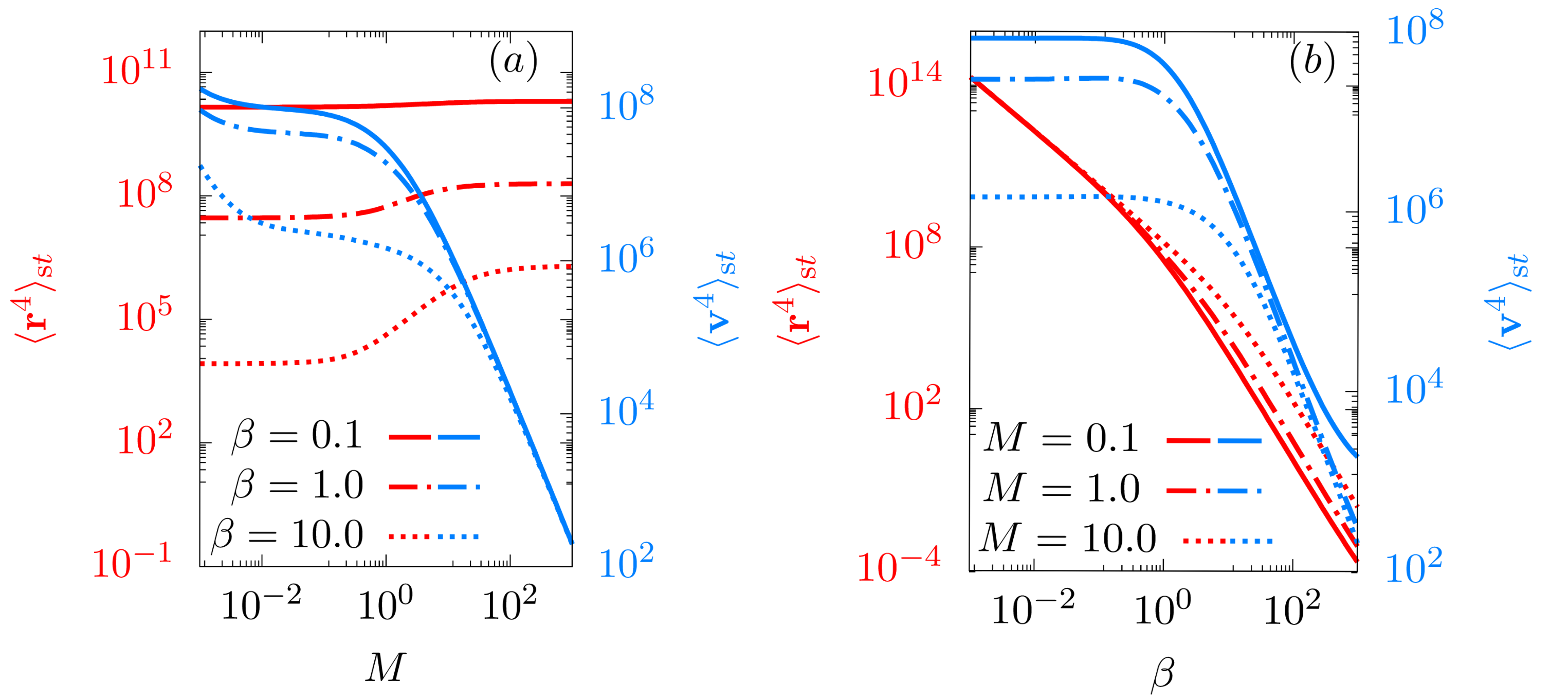}
    \caption{The steady-state of fourth moments $\la \vv^4 \ra_{\rm st}$, $\la \rv^4 \ra_{\rm st}$ as a function of $M$ ($a$) and $\beta$ ($b$) in $2$d. The lines denote plots of equations\,(\ref{eq:v42dst}) and (\ref{eq:r42dst}) at fixed $\Pe = 100$.}
    \label{fig:ss_fourth}
\end{figure}

The steady-state value of the fourth velocity moment in 2d (see \ref{sec:3dresults} for $d=3$ dimension) is given by
\bea
\label{eq:v42dst} \la \vv^4 \ra_{\rm st} = \frac{8}{M^2} + \frac{8}{M(1+\beta+M)}\Pe^2 + \frac{{\cal A}_1}{{\cal B}_1} \Pe^4
\eea 
where ${\cal A}_1 =  \{  \beta \left[ 162 + M \{ 1209+M(2575 + 2M(993+4M(71 + 16M))) \} \right]+ \beta^2 M [297$
$+2M(567+882 M+592 M^2) ]  + 6(1+M)(2+M)(1+4M)(3+4M)+36 \beta^3 M^2(3+8M) \}$ and ${\cal B}_1 = [(1+\beta + M)(3 + 9\beta+M)(1+4M)(2+\beta+4M)(3+4\beta M)(2+M(3+\beta+M))] $.
Figure~\ref{fig:ss_fourth} shows $M$ and $\be$ dependence of $\la \vv^4 \ra_{\rm st}$ at $\Pe=100$, displaying overall decreases of the moment with both the parameters. 
 
Similarly as before, we present a series expansion of $\la \rv^4 \ra (t)$ around $t = 0$ in 2d (see \ref{sec:3dresults} for $d=3$ dimension) with initial values $\rv_0 = \bf{0}$ and $\vv_0 = \bf{0}$. This takes the form
\bea
\label{eq:r42dser}\fl \la \rv^4 \ra (t) = \frac{32}{9M^4} t^6 + \frac{4(M \Pe^2 -4)}{3M^5}t^7 \nn\\
+ \frac{[3232+M \{ -1024 \beta -16(85+16M){\rm Pe }^2 + 45M \Pe^4 \} ]}{720 M^6} t^8 + {\cal O}(t^9).
\eea
The expansion shows $t^6$ scaling at the shortest time, a behavior similar to free iABPs~\cite{Patel_2023}. Subsequently, it shows a crossover to $ \sim t^7$ behavior at $t_{I} = 8M/[3(M\Pe^2 - 4)]$ which is independent of $\be$ but depends on inertia $M$. This behavior is similar to free iABPs. Only at a later time does the impact of trapping show up, as a departure from  $\sim t^7$ sets in at time $t_{II} = 960M(M \Pe^2 -4)/[3232+M \{ -1024 \beta -16(85+16M){\rm Pe }^2 + 45M \Pe^4 \}]$, which depends on trap strength.

At steady state, the fourth moment of displacement in 2d (see \ref{sec:3dresults} for $d=3$ dimension) takes the following form
\bea
 \label{eq:r42dst}\la \rv^4 \ra_{\rm st} = \frac{8}{\beta^2 } + \frac{8(1+M)}{\beta^2 (1+\beta + M)}\Pe^2 + \frac{{\cal A}_2 }{\beta^2 {\cal B}_1}\Pe^4,
\eea
where ${\cal A}_2 = \{ 12(1+M)^2(2+M)(3+M)(1+2M)(1+4M) + 36 \beta^3 M^2 [1+M(2+M)(3+8M)] + \beta^2 M[99+M(750+M(2149+2M(1995+2066M+592M^2)))] + \beta (1+M)(1+4M)[54+M(267+M(524+M(397+2M(83+16M))))] \}$ and ${\cal B}_1$ has the same form as shown after equation~(\ref{eq:v42dst}).
Figure~\ref{fig:ss_fourth} shows $M$ and $\be$ dependence of $\la \rv^4 \ra_{\rm st}$ at $\Pe=100$. With increasing $M$, the moment increases to eventually saturate to $2(2+\Pe^2)^2/\beta^2$ in $2$d. In contrast, it decreases with $\be$.

\subsection{Steady-state Kurtosis and phase diagrams}
The fourth-order moment of a $d$-dimensional Gaussian process $\bm{\phi}$ with $\la \bm{\phi} \ra = \bm{0}$ is given by $\mu_4 = (1+2/d) \la \bm{\phi}^2 \ra^2$. It can be used to define an excess kurtosis of a general stochastic process $\bm{\phi}$ as
\bea
\label{eq:kurtosis}{\cal K}_{\phi} = \frac{\la \bm{\phi}^4 \ra}{\mu_4} -1,
\eea
which vanishes for a Gaussian process and shows non-zero departures for non-Gaussian processes.
We use this definition and steady-state results from the previous subsection to get the kurtosis for velocity and displacement at the steady state.

\subsubsection{Kurtosis in velocity:}
Using $\bm{\phi} = \vv$ in the equation\,(\ref{eq:kurtosis}) and $\la \vv^4 \ra_{\rm st}$ from equation\,(\ref{eq:v42dst}), we get the following form for kurtosis of velocity at steady state in 2d (see \ref{sec:3dresults} for results in $d=3$ dimensions)
\bea
\fl {\cal K}_v = \frac{-M^2 {\rm Pe }^4}{{\cal B}_2[2+2\beta + M(2+\Pe^2)]^2 } \left[ \,\left\{ 6(1+M)(2+M)(1+4M)(2+7M)\right\} \right. \nn\\
\fl \left. +\beta \{54+M(345+M(778 +M(725 +2M(167+84M)))) \} + \beta^2 \{-54+2M(-225 \right. \nn\\
\fl \left. +M(-45+982M+742M^2)) \}  -\beta^3M(99+66M+256M^2)-36 \beta^4 M^2 \right],
\label{eq:kurv2}
\eea
where ${\cal B}_2 = 2(3+9 \beta +M)(1+4M)(2+\beta+4M)(3+4\beta M)(2+M(3+\beta+M))$. 
It is straightforward to check that the limit $\beta \to 0$ of ${\cal K}_v$ gives the known result for free iABPs \cite{Patel_2023}. 

\begin{figure}
    \centering
    \includegraphics[width=12cm]{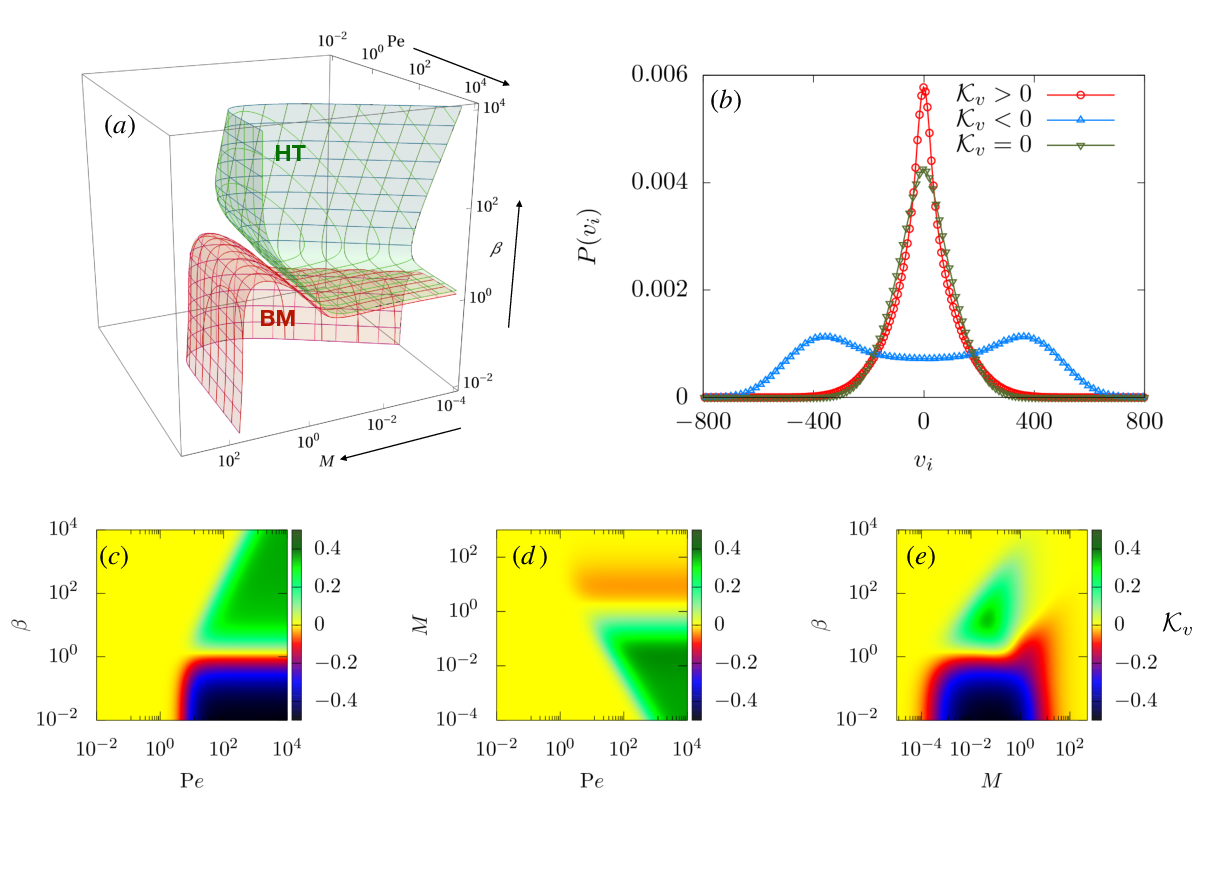}
    \caption{`Phase diagram' and distribution in velocity for 2d iABP: 
    ($a$)~`Phase diagram' as a function of $M, \Pe$, and $\beta$. Regions of bimodal (BM) and heavy-tailed-unimodal (HT) distributions of velocity components are identified using fixed kurtosis surfaces  ${\cal K}_v = -0.03$ (red) and ${\cal K}_v = 0.01$ (green). ($b$)~Marginal distributions of velocity components $P(v_i)$ with  $v_i = v_x$ or $v_y$ obtained from numerical simulations. The parameter value for ${\cal K}_v > 0$ are $M = 0.01$, $\beta = 10$, $\Pe = 500$; for ${\cal K}_v < 0$ are $M = 0.01$, $\beta = 0.1$, $\Pe = 500$; and for ${\cal K}_v = 0$ are $M = 1$, $\beta = 10$, $\Pe = 500$. 
    The `phase diagrams' are further illustrated using heat-maps of ${\cal K}_v$ in $\beta -\Pe$ plane at $M = 0.1$ ($c$), in $M-\Pe$ plane at $\beta = 10$ ($d$), and in $\beta - M $ plane at $\Pe = 100$ ($e$). Plots of kurtosis use equation~(\ref{eq:kurv2}).}
    \label{fig:vphase} 
\end{figure}

At small $\be$, ${\cal K}_v$ starts from ${\cal K}_v = -\{ M^2 \Pe^4 (3 + 7M)\}/[2(3+M)(1+2M)(2+M(2+\Pe^2))^2] $ in 2d, a value same as for free iABPs. 
It varies non-monotonically at intermediate $\beta$ to vanish at large trap strength as ${\cal K}_v \approx \beta^{-2}\,(M^2 \Pe^4 )/(8+32 M)$ in 2d. 
${\cal K}_v$ shows a non-monotonic behavior with $M$ as well and vanishes in the two limits of small and large inertia. 
In the limit of $M \to 0$, it vanishes as ${\cal K}_v \approx M^2\, \{ [-2+3(-1+\be)\be]\Pe^4 \}/[8(1+\be)^2 (2 + \be)(1+3 \be)] $ and in the other limit of $M \to \infty$, it vanishes as ${\cal K}_v \approx -M^{-1}\times 21 \Pe^4/[16  (2+\Pe^2)^2]$ in 2d.
With activity $\Pe$, ${\cal K}_v$ vanishes as $\sim \Pe^4$ in the small activity limit $\Pe \to 0$. With increasing $\Pe$ it shows non-equilibrium departures to saturate to either positive or negative values at high enough $\Pe$.

This velocity kurtosis is used to obtain a ``phase diagram" in the three-dimensional parameter space of activity $\Pe$, trap stiffness $\be$ and inertia $M$~(see figure\,\ref{fig:vphase}($a$)\,). We plot equi-kurtosis planes of positive and negative values, ${\cal K}_v=0.01$~(green plane) and ${\cal K}_v=-0.03$~(red plane) to identify different kinds of non-Gaussian departures. The region above the positive kurtosis plane ${\cal K}_v=0.01$ is denoted HT in the figure to indicate parameter regimes in which heavy-tailed marginal distributions of velocity components are observed, as shown in figure\,\ref{fig:vphase}($b$).  Similarly, the region below the ${\cal K}_v=-0.03$ plane is denoted BM to indicate the bimodal marginal distributions of velocity components for the corresponding parameter values, as shown in figure\,\ref{fig:vphase}($b$). This BM distribution corresponds to an inverted wine bottle or Mexican hat shape for the velocity vector distribution in 2d. To further clarify the phase behavior, we plot two-dimensional projections of the phase diagram using heat-maps of kurtosis values in the $\Pe-\be$, $\Pe-M$, and $M-\be$ planes in figures\,\ref{fig:vphase}($c$)-($e$), respectively. The yellow regions in the figures denote nearly Gaussian distributions of vanishing kurtosis, while the dark blue and green regions stand for negative and positive kurtosis, respectively. A thin region of vanishing kurtosis persists even for large $\Pe$, balanced by $\be$ and $M$. 
At small $\beta$, the kurtosis ${\cal K}_v \leq 0$, as shown in figure~\ref{fig:vphase}($c$) and (e). These figures also show that for sufficiently large $\be$, ${\cal K}_v$ can become positive. 
For a given $\be$ and $\Pe$, it is possible to encounter re-entrant transition from Gaussian to a specific kind of non-Gaussian with positive or negative kurtosis to Gaussian with increasing $M$, as can be seen from figure\,\ref{fig:vphase}($e$).

\subsubsection{Kurtosis in displacement:}
Again, we use equation\,(\ref{eq:kurtosis}) with $\bm{\phi} = \rv$ and $\la \rv^4 \ra_{\rm st}$ from equation\,(\ref{eq:r42dst}) to get the kurtosis of displacement at steady state in 2d (see \ref{sec:3dresults} for results in $d=3$ dimensions)
\bea
\fl {\cal K}_r = \frac{-\beta \Pe^4}{{\cal B}_2 \left[2\beta +(1+M)(2 +\Pe^2)\right]^2} \left[ \left\{ 21(1+M)^3 (2+M)(3+M)(1+2M)(1+4M) \right\} \right. \nn\\
\fl \left. + \beta \left\{ 2(1+M)(27 +M(291+M(1315+M(3061+14M(218+53M))))) \right\} + \beta^2 \left\{ M(99\right. \right.  \nn\\
\fl \left. \left. +M(786  +M(2053+2M(783-128M)))) \right\} -36 \beta^3 M^2\left\{-1+(M-6)M \right\} \right],
\label{eq:kurr2}
\eea
where, ${\cal B}_2$ has the same expression as the one given after equation~(\ref{eq:kurv2}). It is easy to check that the expression for displacement kurtosis reduces to the known results for overdamped ABPs in a harmonic trap~\cite{Chaudhuri_2021} in the limit of $M\to 0$.  

\begin{figure}
    \centering
    \includegraphics[width=12cm]{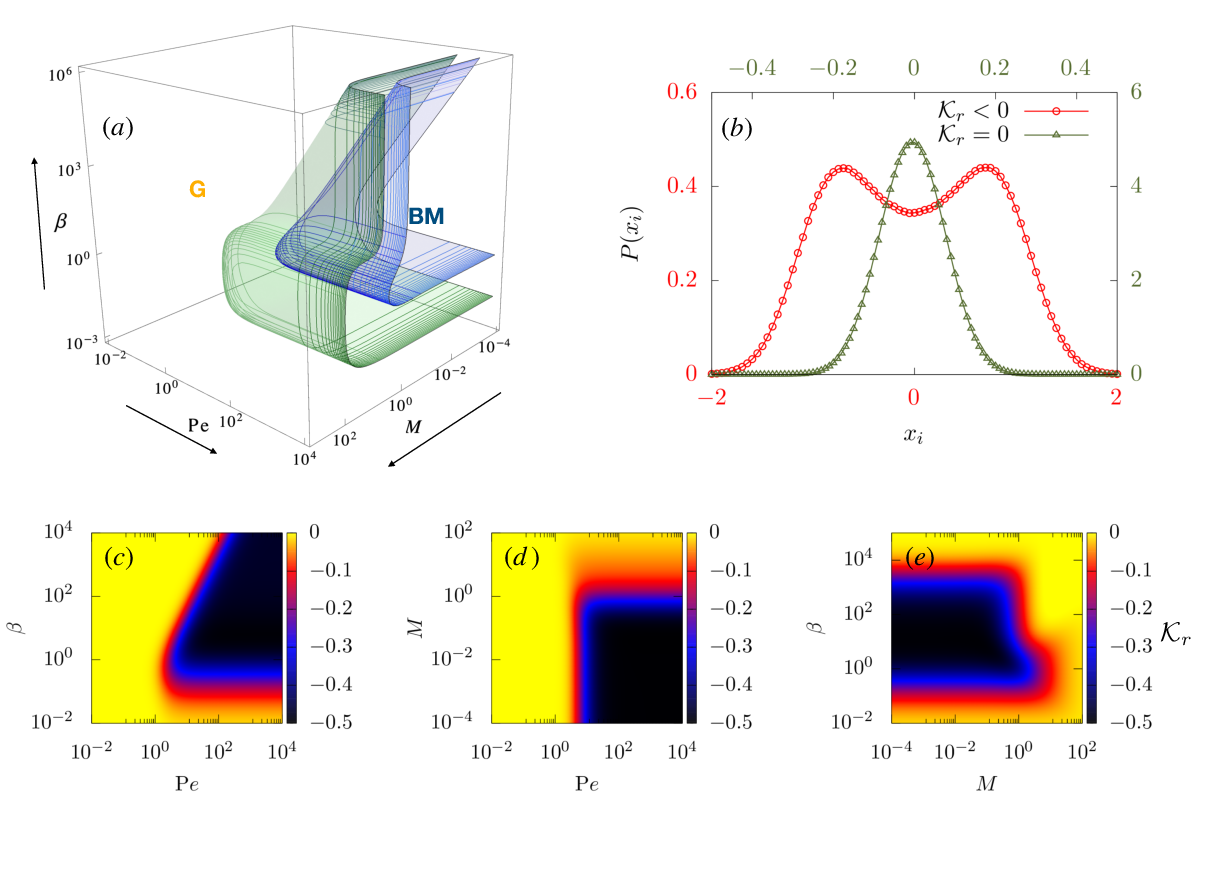}
    \caption{`Phase diagram' and distribution in displacement for 2d iABP: 
    ($a$)~`Phase diagram' as a function of $M, \Pe$, and $\beta$. Regions of Gaussian (G) and bimodal (BM) distributions of displacement components are identified using fixed kurtosis surfaces for ${\cal K}_r = -0.02$ (in green) and ${\cal K}_r = -0.3$ (in blue). ($b$)~The simulated probability distribution of displacements $x_i = x$ or $y$ obtained from numerical simulations. The points are the simulation results, and the lines are guides to the eye. The bimodal distribution at ${\cal K}_r < 0$ corresponds to $M = 0.1$, $\beta = 10$, $\Pe = 10$ and the Gaussian distribution at ${\cal K}_r = 0$ corresponds to $M = 0.1$, $\beta = 200$, $\Pe = 10$.  ($c$) -- ($e$) shows heat-maps of kurtosis ${\cal K}_r$ across three planar sections of ($a$), in $\beta -\Pe$ plane with $M = 0.1$ ($c$), in $M-\Pe$ plane with $\beta = 10$ ($d$), and in $\beta - M $ plane with $\Pe = 100$ ($e$). Plots of kurtosis ${\cal K}_r$ uses equation~(\ref{eq:kurr2}). }
    \label{fig:rphase}
\end{figure}

For vanishing $M$, kurtosis of displacement saturates to ${\cal K}_r = -\{\be \Pe^4 (7 + 3 \be) \}/[2(2+\be)(1+3 \be)(2+2\be + \Pe^2)^2] $ in 2d, a negative value that describes overdamped ABPs in a harmonic trap~\cite{Chaudhuri_2021}. 
It increases monotonically with $M$ to vanish as ${\cal K}_r \approx -M^{-1} \times 21 \Pe^4/[16  (2 + \Pe^2)^2]$ in the limit $M \to \infty$.
With activity, ${\cal K}_r$ vanishes as ${\cal K}_r \sim -\Pe^4$ as $\Pe$ vaishes. It becomes negative with increasing activity to saturate to $M$ and $\be$ dependent values.
With trap strength, ${\cal K}_r$ vanishes in the two limits of small and large $\beta$. In the limit $\be \to 0$, it vanishes linearly as ${\cal K}_r \approx -\be \times 7 \Pe^4 / [4(2+\Pe^2)^2]$. On the other hand, in the limit $\beta \to \infty$, it vanishes as $\be^{-2} \times [(-1+M(M-6)) \Pe^4]/[(8+32M)]$.

The displacement kurtosis is used to obtain a ``phase diagram" in the three-dimensional parameter space of activity $\Pe$, trap stiffness $\be$ and inertia $M$~(see figure\,\ref{fig:rphase}($a$)\,). Here, we plot equi-kurtosis planes of two different negative values, ${\cal K}_r=-0.02$~(green plane) and ${\cal K}_r=-0.3$~(blue plane) to identify different degrees of non-Gaussian departures. Note that unlike ${\cal K}_v$, ${\cal K}_r$ is never positive. The region to the left of the negative kurtosis plane ${\cal K}_r=-0.02$ is denoted G in the figure to indicate parameter regimes in which  Gaussian distributions of displacement of vanishing kurtosis are observed, as shown in figure\,\ref{fig:rphase}($b$). Similarly, the region to the right of the ${\cal K}_r=-0.3$ plane is denoted BM to indicate the bimodal marginal distributions of displacement components for the corresponding parameter values, as shown in figure\,\ref{fig:rphase}($b$). To further clarify the phase behavior, we plot three two-dimensional projections of the phase diagram using heat-maps of kurtosis values in the $\Pe-\be$, $\Pe-M$, and $M-\be$ planes in figures\,\ref{fig:rphase}($c$)-($e$), respectively. The yellow regions in the figures denote nearly Gaussian distributions of vanishing kurtosis, while the dark blue regions stand for negative kurtosis. At a given $M$, a re-entrant transition from Gaussian to BM to Gaussian can be obtained by increasing $\be$, as can be seen in figures\,\ref{fig:rphase}($c$) and ($e$). While $\Pe$ drives the displacement to non-equilibrium BM distributions, large $M$ moves it back towards equilibrium-like Gaussian distributions. 

The position phase diagrams also suggest a method for particle segregation in active matter using a trap. The differential accumulation of iABPs with small activity near the trap center and that with high activity at the {\em boundary} can be utilized to segregate these two kinds of particles. More interesting is the inertia dependence. In the presence of intermediate trap strength, particles with large $M$ accumulate near the trap center, and those with small $M$ near the trap {\em boundary}, allowing for a method of segregating massive and lighter iABPs.


\label{sec:conclusio}
\section{Outlook}
\label{sec_conclusion}
In conclusion, we presented an exact method to calculate all the time-dependent dynamical moments of $d$-dimensional iABPs in a harmonic trap. We derived the time evolution of several dynamical moments, including the second and fourth moments of velocity and displacement, to observe excellent agreement with numerical simulations. 
Remarkably, the steady-state kinetic temperature and an estimate of effective diffusivity depend on inertia and trap stiffness, with inertial dependence of the diffusivity entering through the potential. This suggests a possible inertia-dependence of diffusivity in a system of interacting iABPs, as has been observed in recent numerical simulations~\cite{khali2023does}.

While the non-equilibrium activity drives the steady-state distributions of velocity and displacement away from the equilibrium Gaussian, large inertia and trap strength suppress non-Gaussian fluctuations to bring back equilibrium-like features. These behaviors are quantified using the exact expressions for excess kurtosis in velocity and displacement. We presented detailed ‘phase diagrams’ using them for three control parameters: inertia, trap strength, and activity. The position phase diagram suggests a method using harmonic traps to segregate particles based on their activity and inertia.  

The excess velocity kurtosis showed both negative and positive departures associated with distributions of velocity components being bimodal and heavy-tailed-unimodal, respectively. Further, the velocity kurtosis displays non-monotonic variation with $M$,  indicating the possibility of a `re-entrant transition' from Gaussian to non-Gaussian to Gaussian.
Our exact expression for kurtosis of displacement revealed a qualitatively different `phase diagram'. It showed two `phases', one with a negative value associated with the ‘active’ bimodal distribution corresponding to particle accumulation near the {\em boundary} and the other with a vanishing value for the ‘passive’ Gaussian phase corresponding to predominant particle accumulation near trap center. A non-monotonicity and re-entrant `transition' is observed for displacement kurtosis as well, with increasing $\be$ but not with $M$.         

 \begin{table}
\begin{center}
\begin{tabular}{ | m{6em} | m{4.5em} |m{4.5em} |m{6.5em} |m{3.5em} |m{4.em} | m{6.0em} | }
  \hline
 Experiment & $m (g)$ & $v_0 (cm/s)$ & $\gamma \equiv \frac{1}{\mu} (g/s) $ & $k(g/s^2)$ & $D(cm^2/s)$ & $D_r(s^{-1})$ \\ 
  \hline
 vibrobot~\cite{Scholz_2018} & $0.33-4.07 $ & $5.7-9.3$ & $\sim 1.65-26.3$ & $-$ & $0.36-2.2$ & $0.59-2.7$\\ 
  \hline
  hexbugs~\cite{dauchot_2019} & $7.5 $ & $10$ & $\sim 10.9-54.8$ & $ \sim 80.2$ & $-$ & $\sim 0.006-0.06$\\ 
  \hline
  hexbugs~\cite{ignacio_2021} & $7.2 $ & $13$ & $\sim 720$ & $90$ & $-$ & $0.3$\\ 
   \hline
\end{tabular}
\caption{Estimate of experimentally achieved parameter range. 
}
\label{table2}
\end{center}
\end{table}

The analytical predictions in our study are amenable to direct experimental verifications. For example, we consider recent experiments on vibrobots~\cite{Scholz_2018} and hexbugs~\cite{dauchot_2019} to extract relevant parameter values and compare them with the dimensionless parameter regimes used in this paper. 
Some of these experiments used the parabolic dish to generate an effective harmonic trapping~\cite{dauchot_2019,ignacio_2021}, where
the trap-strength $k$ is determined by the geometry of the dish and gravity, using $kr^2/2 = mgh$ where $m$ is the mass of the active particle, $g$ is the gravitational acceleration, and $h$ is the maximum height of the dish of radius $r$. 
We list the parameter values estimated from Ref.~\cite{Scholz_2018,ignacio_2021,dauchot_2019} in Table-\ref{table2}.
Wherever a required measure is not available explicitly, we use the values supplied in other related experiments; e.g., for effective translational diffusivity, we use $D \sim 1 \,{\rm cm}^2/{\rm s}$ for both vibrobots and hexbugs~(see Table-\ref{table3}). This is a value within the range measured in Ref.~\cite{Scholz_2018}.   Using this approach, we report the estimated dimensionless parameters in Table-\ref{table3}.
The numbers realized, $M \sim 0.001 - 20$, $\Pe\sim 1 - 130$, and $\be\sim 0.1 - 250$, span a broad enough range to cover approximately all the different velocity and displacement phases we predicted. Further, the parameter values used in figures\,\ref{fig:firstmoments},\,\ref{fig:secondorder},\,\ref{fig:fourthorder} lie within the range estimated in this table. Similar experiments can be used to test our predictions for transitions between active and passive behaviors.  

\begin{table}
    \begin{tabular}{|m{8em} | m{8 em} |m{8 em} |m{8 em} |}
         \hline
 Experiment & $M$ & $\Pe$ & $\beta $  \\ 
 \hline
  vibrobot~\cite{Scholz_2018} & $\sim 1.77 - 18.5 $ & $\sim 4.01-12.5$ & $-$  \\
 \hline
 hexbugs~\cite{dauchot_2019} & $\sim 0.0008 - 0.04 $ & $\sim 40.8 - 129$ & $\sim 24.34 - 243.4$ \\
 \hline
 hexbugs~\cite{ignacio_2021} & $\sim 0.003 $ & $\sim 23$ & $\sim 0.42$ \\
 \hline
    \end{tabular}
    \caption{Estimate of dimensionless parameter values obtained from experiments. For hexbugs of similar sizes as vibrobots, we use $D = 1\,\rm{cm}^2/\rm{s}$, picked from the range of measured vibrobot diffusivities~\cite{Scholz_2018}. }
    \label{table3}
\end{table}

\section*{Acknowledgments}
DC thanks Abhishek Dhar and Fernando Peruani for collaborations on related topics, acknowledges research grants from DAE (1603/2/2020/IoP/R\&D-II/150288) and SERB, India (MTR/2019/000750), and thanks ICTS-TIFR, Bangalore, for an Associateship.

\appendix

 \section {Fourth moment of velocity and position} \label{sec:fourthorder_appendix}
The calculation of $\la \vv^4 \ra_s$ and $\la \rv^4 \ra_s$ utilises equation (\ref{eq:observable}) and follows the following steps:
\bea
\label{eq:r4s}&&\fl (i)\, s \la \rv^4 \ra_s = \rv_0^4 + 4 \la (\vv \cdot \rv)\rv^2 \ra_s, \\
&&\fl (ii)\, (s+1/M) \la (\vv \cdot \rv) \rv^2 \ra_s =  (\vv_0 \cdot \rv_0) \rv_0^2  + \la \vv^2 \rv^2 \ra_s + 2 \la (\vv \cdot \rv)^2 \ra_s + \Pe\la (\uv \cdot \rv) \rv^2 \ra_s/M \nn\\
&& \fl - \beta \la \rv^4 \ra_s/M, \\
&& \fl (iii) \,(s+2/M) \la \vv^2 \rv^2 \ra_s =  \vv_0^2 \rv_0^2  + 2 \la (\vv \cdot \rv) \vv^2 \ra_s + 2 \Pe\la (\uv \cdot \vv) \rv^2 \ra_s/M - 2\beta \la(\vv \cdot \rv) \rv^2 \ra_s /M
\nn\\
&& \fl + 2d\la \rv^2\ra_s/M^2, \\
&& \fl (iv)\,(s+3/M) \la (\vv \cdot \rv) \vv^2 \ra_s =   (\vv_0 \cdot \rv_0)\vv_0^2  + \la \vv^4 \ra_s + \Pe\la (\uv \cdot \rv)\vv^2 \ra_s/M 
+ 2\Pe\la(\uv \cdot \vv)(\vv \cdot \rv) \ra_s /M
\nn\\ 
&& \fl - \beta \la \vv^2 \rv^2 \ra_s/M - 2 \beta\la(\vv \cdot \rv)^2 \ra_s/M 
+2(d+2)\la \vv \cdot \rv \ra_s/M^2, \\
&& \fl (v) \, [s+2/M+(d-1)] \la (\uv \cdot \vv)(\vv \cdot \rv) \ra_s = (\uv_0 \cdot \vv_0)(\vv_0 \cdot \rv_0)  + \la (\uv \cdot \vv) \vv^2 \ra_s + \Pe\la \vv \cdot \rv \ra_s/M \nn\\
&& \fl + \Pe \la(\uv \cdot \vv)(\uv \cdot \rv) \ra_s/M - \beta \la (\uv \cdot \rv)(\vv \cdot \rv) \ra_s/M - \beta \la (\uv \cdot \vv) \rv^2 \ra_s/M + 2\la \uv \cdot \rv \ra_s/M^2, \\
&& \fl (vi)\,[s+3/M +(d-1)] \la ( \uv \cdot \vv) \vv^2 \ra_s = (\uv_0 \cdot \vv_0) \vv_0^2  + \Pe\la \vv^2 \ra_s/M + 2\Pe\la(\uv \cdot \vv)^2 \ra_s/M \nn\\
&& \fl - \beta \la(\uv \cdot \rv)\vv^2 \ra_s/M - 2 \beta \la (\uv \cdot \vv)(\vv \cdot \rv) \ra_s/M + 2(d+2)\la(\uv \cdot \vv) \ra_s/M^2, \\
&& \fl (vii)\,[s+2/M+2d] \la(\uv \cdot \vv)^2 \ra_s = (\uv_0 \cdot \vv_0)^2   + 2 \Pe\la (\uv \cdot \vv) \ra_s/M- 2 \beta \la (\uv \cdot \vv)(\uv \cdot \rv) \ra_s/M \nn\\
&&\fl +2/(sM^2)+ 2 \la \vv^2 \ra_s,     \\
&& \fl (viii) \,[s+2/M +(d-1)] \la (\uv \cdot \rv) \vv^2 \ra_s = (\uv_0 \cdot \rv_0) \vv_0^2  + \la (\uv \cdot \vv) \vv^2 \ra_s + 2 \Pe\la(\uv \cdot \vv)(\uv \cdot \rv) \ra_s/M\nn\\
&& \fl - 2 \beta \la(\uv \cdot \rv)(\vv \cdot \rv) \ra_s/M + 2d \la (\uv \cdot \rv) \ra_s/M^2,  \\
&& \fl (ix)  \,[s+1/M+2d] \la (\uv \cdot \vv)(\uv \cdot \rv) \ra_s =  (\uv_0 \cdot \vv_0)(\uv_0 \cdot \rv_0)  +\la (\uv \cdot \vv)^2 \ra_s + \Pe\la(\uv \cdot \rv) \ra_s/M \nn\\ 
&& \fl - \beta \la (\uv \cdot \rv)^2 \ra_s/M +2 \la (\vv \cdot \rv) \ra_s, \\
&& \fl (x) \, [s+1/M+(d-1)] \la (\uv \cdot \vv) \rv^2 \ra_s =  (\uv_0 \cdot \vv_0) \rv_0^2  + 2 \la (\uv \cdot \vv)(\vv \cdot \rv) \ra_s + \Pe\la \rv^2 \ra_s/M \nn\\ 
&& \fl - \beta \la(\uv \cdot \rv) \rv^2 \ra_s/M, \\
&& \fl (xi)\,(s+2/M) \la (\vv \cdot \rv)^2 \ra_s =  (\vv_0 \cdot \rv_0)^2  + 2 \la (\vv \cdot \rv)\vv^2 \ra_s + 2\Pe\la (\uv \cdot \rv)(\vv \cdot \rv)\ra_s/M \nn\\
&& \fl - 2 \beta \la (\vv \cdot \rv) \rv^2 \ra_s/M + 2 \la \rv^2 \ra_s/M^2, \\
&& \fl (xii) \, [s+(d-1)] \la (\uv \cdot \rv) \rv^2 \ra_s =  (\uv_0 \cdot \rv_0) \rv_0^2  + \la (\uv \cdot \vv) \rv^2 \ra_s + 2 \la (\uv \cdot \rv)(\vv \cdot \rv) \ra_s, \\
&& \fl (xiii) \, (s+4/M) \la \vv^4 \ra_s =  \vv_0^4  + 4 \Pe\la (\uv \cdot \vv) \vv^2 \ra_s/M - 4\beta \la (\vv \cdot \rv) \vv^2 \ra_s/M \nn\\
&& \fl + 4(d+2) \la \vv^2 \ra_s/M, \\
&& \fl (xiv) \,[s+1/M+(d-1)] \la (\uv \cdot \rv)(\vv \cdot \rv) \ra_s =  (\uv_0 \cdot \rv_0)(\vv_0 \cdot \rv_0)  + \la (\uv \cdot \vv)( \vv \cdot \rv) \ra_s + \nn\\
&& \fl \la ( \uv \cdot \rv) \vv^2 \ra_s + \Pe\la(\uv \cdot \rv)^2 \ra_s/M - \beta \la  (\uv \cdot \rv) \rv^2 \ra_s/M, \\
\label{eq:ur2s}&& \fl (xv) \, (s+2d) \la (\uv \cdot \rv)^2 \ra_s =  (\uv_0 \cdot \rv_0)^2  + 2 \la (\uv \cdot \vv)(\uv \cdot \rv) \ra_s +2 \la \rv^2 \ra_s. 
\eea
where $\la \uv \cdot \rv \ra_s = \la r_{\parallel} \ra_s$, $ \la \vv \cdot \rv \ra_s$, $\la \uv \cdot \vv \ra_s = \la v_{\parallel} \ra_s$, $\la \rv^2 \ra_s$, and $\la \vv^2 \ra_s$ were already calculate in equations\,(\ref{eq:urs}),~(\ref{eq:vrs}),~(\ref{eq:uvs}),~(\ref{eq:r2s}), and~($\ref{eq:v2s}$) respectively. Solving these coupled equations, one can get the fourth-order moments in the Laplace space such as $\la \rv^4 \ra_s$ and $\la \vv^4 \ra_s$ and the inverse Laplace transform can be used to get the full-time evolution.


\section{Kurtosis and fourth moments in $d = 3$ dimensions} \label{sec:3dresults}
Solving equations\,(\ref{eq:r4s}) $-$ (\ref{eq:ur2s}) for $\la \rv^4 \ra_s$ and $\la \vv^4 \ra_s$ and taking the inverse Laplace transform, we get the full-time evolution in arbitrary dimension.
We expand these expressions around $t=0$ for initial values $ \rv_0  = \bf{0}$ and $ \vv_0 = \bf{0}$ in $d=3$ dimensions to get 
\bea
\label{eq:v43dser}\fl \la \vv^4 \ra (t) = \frac{60}{M^4} t^2 + \frac{20(M \Pe^2 -6)}{M^5} t^3 \nn\\
+ \frac{[420 + M \{ -120 \beta -40 (M + 3)\Pe^2 + 3M \Pe^4 \}]}{3M^6}t^4 + {\cal O} (t^5), \\
\label{eq:r43dser} \fl \la \rv^4 \ra (t) = \frac{20}{3M^4}t^6 + \frac{5(M\Pe^2 -6)}{3M^5} t^7 \nn\\
+ \frac{[1212+M \{ -384 \beta -4(32 M + 85)\Pe^2 + 9M \Pe^4 \}]}{144 M^6} t^8 + {\cal O}(t^9).
\eea
The comparison of equations\,(\ref{eq:v43dser}) and (\ref{eq:r43dser}) with the equations\,(\ref{eq:v42dser}) and (\ref{eq:r42dser}) respectively suggests that the nature of crossover in the dynamic is independent of dimensions. However, the crossover time itself depends on the dimensions.

In the asymptotic limit of a long time, we get the steady-state results in $d=3$ dimensions as 
\bea
\la \vv^4 \ra_{\rm st} = \frac{15}{M^2} + \frac{20}{M(2 + \beta + 4M)} \Pe^2 + \frac{4\,{\cal A}_3}{{\cal B}_3} \Pe^4,
\eea
where ${\cal A}_3 = \{ 18(1+M)(1+2M)(3+5M)(1+6M) + 3\beta [36+M(423+M(1495 +2M(1021+12M(49+20M))))] + \beta^2M[198 + M(1371+50M(67+68M))] + 72 \beta^3 M^2 (1+5M) \} $ and ${\cal B}_3 = \{ (2 + \beta + 4M) (6 + 9\beta + 4M) (1+6M)(3 + \beta + 9M) (3 + 4\beta M) (2+M(6 + \beta + 4M))\}$.

Further, 
\bea
\la \rv^4 \ra_{\rm st} = \frac{15}{\beta^2} + \frac{10(2M +1)}{\beta^2 (2+ \beta + 4M)} \Pe^2 + \frac{{\cal A}_4}{\beta^2 {\cal B}_3} \Pe^4,
\eea
where ${\cal A}_4 = \{ 30(1+M)(1+2M)^2 (3+2M)(1+3M)(1+6M) + \beta [(1+2M)(1+6M)(54+M(435 +M(1325+2M(885 +4M(167+60M)))))] + \beta^2 M [99+2M(633+10M(311+M(881+10M(135+68M))))] + 36 \beta^3 M^2 [1+2M(5+4M(6+5M))] \} $. The limiting behavior of $\la \vv^4 \ra_{\rm st}$ and $\la \rv^4 \ra_{\rm st}$ with $M$ and $\beta$ in $d = 3$ dimensions shows the same qualitative feature as in 2d. 

Following the definition of kurtosis from equation\,(\ref{eq:kurtosis}) and the steady state results from this section for $d = 3$ dimensions, we write the kurtosis of velocity and displacement in $d=3$ dimensions as  
\bea
\fl {\cal K}_v = \frac{-4M^2 \Pe^4}{{\cal B}_4} \left[ \left\{ 72(1+M)(1+2M)(1+6M)(3+11M) \right\} + \beta \left\{12(15+M(151+4M(139 \right. \right. \nn\\
\fl \left. \left. +M(230+M(215+198M)))))  \right\} + \beta^2 \left\{ -54+2M(-375+2M(307+4110M+5324M^2)) \right\} \right. \nn\\ 
\left. \fl + \beta^3  M(-99+M(165-214M))  -36 M^2 \beta^4 \right],
\eea
where ${\cal B}_4 =5(6+9\beta + 4M)(1+6M)(3+\beta+9M)(3+4\beta M)(2+M(6+\beta+4M))(6+3 \beta+2M(6+ \Pe^2))^2$,

and
\bea
\fl {\cal K}_r = \frac{-2 \beta \Pe^4}{{\cal B}_5} \left[ \left\{ 66(1+M)(1+2M)^3(3+2M)(1+3M)(1+6M)\right\} + \beta \left\{ (1+2M)(54 \right. \right. \nn\\
\fl \left. \left. +M(957+M(7187+2M(13865+88M(266+121M))))) \right\} + \beta^2 \left\{ M(99+2M(669 \right. \right. \nn\\
\fl \left. \left. +M(3107+M(4313-214M)))) \right\} -36 M^2 \beta^3 \left\{-1+2M(M-5)  \right\} \right],
\eea
where ${\cal B}_5 =5(6+9\beta + 4M)(1+6M)(3+\beta+9M)(3+4\beta M)(2+M(6+\beta+4M))[3 \beta+(1+2M)(6+ \Pe^2)]^2 $. 

The kurtosis of velocity and position shows the same qualitative features in 3 dimensions as in for 2d; however, the \textit{boundary} separating the different `phases' gets shifted.

\section*{References}

\bibliographystyle{unsrt} 


\end{document}